\documentclass{article}
\usepackage{spconf,amsmath,graphicx}
\usepackage{enumitem}
\usepackage{cite}
\usepackage{amsfonts,amssymb,amsthm,bm,bbm} %my math packages
\usepackage[caption=false,font=footnotesize]{subfig}
\usepackage{epsfig}
\usepackage{epstopdf}

\renewcommand{\vec}[1]{\mathbf{#1}} %command for bold vector
\newtheorem{proposition}{Proposition} % proposition environment

\usepackage{flushend} %to balance last page

\title{Downlink Channel Spatial Covariance Estimation in Realistic FDD Massive MIMO Systems}
\name{Lorenzo Miretti$^{\star}$ \qquad Renato L.G. Cavalcante$^{\dagger}$ \qquad Slawomir Sta\'nczak$^{\dagger}$}

\address{$^{\star}$ EURECOM  \\
    $^{\dagger}$ Fraunhofer Heinrich Hertz Institute and Technical University of Berlin}
		
\begin{document}
\ninept
\setlength{\belowdisplayskip}{1.2pt}
\setlength{\belowdisplayshortskip}{1.2pt}
\setlength{\abovedisplayskip}{1.2pt}
\setlength{\abovedisplayshortskip}{1.2pt}
 \setlength{\belowcaptionskip}{-40pt}

\maketitle
\begin{abstract}
The knowledge of the downlink (DL) channel spatial covariance matrix at the BS is of fundamental importance for large-scale array systems operating in frequency division duplexing (FDD) mode. In particular, this knowledge plays a key role in the DL channel state information (CSI) acquisition. In the massive MIMO regime, traditional schemes based on DL pilots are severely limited by the covariance feedback and the DL training overhead. To overcome this problem, many authors have proposed to obtain an estimate of the DL spatial covariance based on uplink (UL) measurements. However, many of these approaches rely on simple channel models, and they are difficult to extend to more complex models that take into account important effects of propagation in 3D environments and of dual-polarized antenna arrays. In this study we propose a novel technique that takes into account the aforementioned effects, in compliance with the requirements of modern 4G and 5G system designs. Numerical simulations show the effectiveness of our approach.
\end{abstract}
\section{Introduction}
In this study we propose a technique to estimate the downlink (DL) channel spatial covariance matrix $\vec{R}^d$ in realistic massive MIMO systems operating in frequency division duplexing (FDD) mode. Although typical massive MIMO systems operate in time division duplexing (TDD) mode, the extension of this technology to FDD mode is of great practical interest \cite{bjornson2016massive} \cite[Chapter~8.4]{marzetta2016fundamentals}. The capability of the base station (BS) to access accurate and efficient $\vec{R}^d$ estimates has emerged as an enabling technology to address practical implementation issues of FDD large-scale array systems \cite{scalingMIMO,Caire1}, as it provides long-term information that is essential for beamforming and for CSI acquisition \cite{Caire2,Cottat,akdeniz2014millimeter,corr_MMSE,CorrMIMO}. 

Conventional FDD systems typically acquire $\vec{R}^d$ by using traditional DL training and uplink (UL) covariance feedback schemes. However, in massive MIMO systems, due to the large size of the covariance matrices and to the large DL training overhead, the traditional schemes become unfeasible. To overcome the drawbacks of conventional systems, in \cite{mypaper}, we propose a scheme to infer $\vec{R}^d$ from the observed UL covariance $\vec{R}^u$. This approach is based on projection methods and has many benefits. In particular, it eliminates the continuous DL training and covariance feedback loop required by conventional direct $\vec{R}^d$ estimation techniques. Moreover, it is completely transparent to the user equipment (UE), hence it can be implemented in compliance with current standards.  %On top of the overhead reduction, an additional benefit is that operators can immediately apply the proposed scheme to boost the already implemented beamforming and CSI acquisition algorithms in compliance with current standards, because the proposed mechanism is completely transparent to the user equipment (UE). 

Related state-of-the-art techniques for UL to DL covariance matrix conversion in FDD systems have been considered in \cite{FC,Splines,caireCov,Dict1}. The main limitation of \cite{mypaper} and of the related works \cite{FC,Splines,caireCov} is that they are based on simple channel models that do not meet the requirements of modern 4G and 5G system designs. More precisely, the approaches in \cite{FC,Splines,caireCov} seem hard to generalize to arrays with arbitrary geometries and non-isotropic antennas. Furthermore, \cite{mypaper,FC,caireCov,Splines} do not consider propagation effects of 3D environments and, most importantly, dual-polarized antenna arrays. In contrast, the technique proposed in \cite{Dict1} is able to cope with these design requirements. However, it is a machine learning approach that relies on the acquisition of a training set, and hence it is significantly more complex.

In this study, we propose a simple training-free approach that takes into account the aforementioned effects. To this end, in Sect. \ref{sec:model} we present a realistic multipath channel model, and we derive expressions for $\vec{R}^d$ and $\vec{R}^u$ under the assumption of both narrow-band and wide-band OFDM systems. Then, in Sect. \ref{sec:main_algo}, we describe the proposed scheme which infers $\vec{R}^d$ from $\vec{R}^u$ by exploiting the proposed covariance model. The resulting algorithm is based on the joint estimation of the two angular power spectra for the vertical (V-APS) and for the horizontal (H-APS) polarization (defined in Sect. \ref{sec:model}). The key idea behind our approach is the definition of a suitable Hilbert space that allows us to formalize the joint V-APS and H-APS estimation problem as a convex feasibility problem. This enables us to adopt standard projection-based solutions inspired by \cite{mypaper}. Furthermore, in Sect. \ref{sec:UPA} we provide implementation details for the case of a cross-polarized uniform planar array (UPA) at the BS. Finally, in Sect. \ref{sec:sim} we evaluate the proposed approach by means of numerical simulations.

\textit{Notation:} We use boldface to denote vectors and matrices.  $( \cdot)^T$ and $( \cdot )^H$ denote respectively the transpose and Hermitian transpose. By defining the set $I \subset \mathbb{R}^2$, $C^0[I]$ and  $L^2[I]$ denote, respectively, the set of all continuous functions and the set of all square Lebesgue integrable functions over $I$. $\Re[\cdot]$ and $\Im[\cdot]$ denote respectively the real and the imaginary parts. We denote the imaginary unit by $j$. Throughout the paper, superscripts $(\cdot)^u$ and $(\cdot)^d$ indicate respectively UL and DL matrices, vectors, or functions when we need to emphasize the dependency on the carrier frequency.

\section{System Model}
\label{sec:model}

We consider a MU-MIMO channel between a BS with $N \gg 1$ antennas and single-antenna UEs. We denote by $\vec{h}$ the channel vector of an arbitrary user. In the remainder of this section we describe the underlying models that we use for designing our scheme. First, in Sect. \ref{ssec:realistic_multipath_model} we review a widely-considered narrow-band multipath model that takes into account 3D propagation and polarization effects. Then, in Sect. \ref{ssec:covariance_expressions} we present analytical expressions for $\vec{R}^d$ and $\vec{R}^u$ based on the considered channel model. Finally, in Sect. \ref{ssec:OFDM_expression} we obtain analogous expressions also for wide-band OFDM systems. Due to the space limitation, all proofs of this section are omitted.

%\subsection{Second-order Statistics Properties}
%By dropping the UL and DL superscripts for brevity, we assume $\vec{h}[m]$ to be a random process correlated both in time and in the spatial domain, with spatio-temporal covariance matrix
%\begin{equation*}
%    \vec{R}_h[m,m+n] := \mathbb{E}\left[\vec{h}[m]\vec{h}[m+n]^H\right].
%\end{equation*}
%We assume $\vec{R}_h$ to be slowly-varying with $m$, such that, for a time window of length $T_{\text{WSS}}$, we can approximate the channel as a wide-sense stationary (WSS) process, i.e. with spatio-temporal covariance matrix  $\vec{R}_h[n]$ that depends only on the lag $n$. In this work we focus on the spatial covariance matrix $\vec{R}:= \vec{R}_h[0]$. Furthermore, we define the coherence block interval $T_c$ as the minimum lag $n$ such that $\vec{h}[m]$ and $\vec{h}[m+n]$ can be considered independent. Moreover, we assume $T_{\text{WSS}} \gg T_c$, which is a reasonable assumption in usual scenarios \cite{Cottat,Caire3}.

\subsection{Realistic Directional Multi-path Model}
\label{ssec:realistic_multipath_model}

In this section we consider an extension of classical 2D directional multi-path channel models (e.g., the ones adopted in \cite{Cottat} and \cite{Caire3}) that take into account 3D propagation and polarization effects. With this extension, we show later in Sect. \ref{sec:main_algo} that we are able to address the UL-DL covariance conversion problem by using projection methods in a Hilbert space different from that in \cite{mypaper}. In more detail, by dropping the frequency dependent superscript for simplicity, we model the channel vector $\vec{h}$ at an arbitrary time $t = t_0$ according to the 3GPP narrow-band clustered directional multi-path model \cite[Eq.~(7.3-22)]{3GPP-3D}. In this model, we have $\vec{h} = \sum_{c=1}^{N_c}\vec{h}_c$, where
\begin{equation*}
    \vec{h}_c := \sqrt{\dfrac{\alpha_c}{N_p}}\sum_{i=1}^{N_p}\vec{A}(\bm{\theta}_{ic})\begin{bmatrix}
        e^{j\varphi_{VV,ic}}     & \dfrac{e^{j\varphi_{VH,ic}} }{\sqrt{K_{ic}}}      \\
         \dfrac{e^{j\varphi_{HV,ic}}}{\sqrt{K_{ic}}}     & e^{j\varphi_{HH,ic}}       \\
\end{bmatrix}\vec{B}(\bm{\phi}_{ic})^H.
\end{equation*}
The notation used here is defined as follows:
\begin{itemize}[topsep=0pt,parsep=0pt,partopsep=0pt,leftmargin=10pt,labelwidth=6pt,labelsep=4pt]
\item $N_c \in \mathbb{N}$ denotes the number of clusters of scatterers, and $N_p\in \mathbb{N}$ denotes the associated number of subpaths. This terminology derives from the classical geometry-based stochastic channel model (GSCM) \cite[Chapter~7]{MolishB}.
\item $\bm{\theta}_{ic}\in \mathbb{R}$ and $\bm{\phi}_{ic}\in \mathbb{R}$ are, respectively, either the direction of departure (DoD) and of arrival (DoA) of subpath $i$ of cluster $c$ for the DL case, or the DoA and DoD of subpath $i$ of cluster $c$ for the UL case. The directions $\bm{\theta}_{ic}$ and $\bm{\phi}_{ic}$ are defined as tuples taking values in the set $\Omega := [-\pi, \pi]\times[0,\pi]$, which represents the azimuth and the zenith of a spherical coordinate system. They are drawn independently from a continuous joint distribution $f_c(\bm{\theta},\bm{\phi})$, and they are assumed to be equal for UL and DL. This DoD/DoA statistical modeling approach, which is very popular in the literature \cite{Caire3,Cottat,MolishB}, generalizes the model given by 3GPP \cite{3GPP-3D}, where only the main cluster angles are random and the subpaths angles are obtained from tables. 
\item $\vec{A}: \Omega \to \mathbb{C}^{N\times2}$ is the dual polarized antenna array response of the BS. In FDD systems, $\vec{A}^d$ is different from $\vec{A}^u$. The columns of $\vec{A}$ are denoted by $[\vec{a}_V,\vec{a}_H] := \vec{A}$, and they represent the array responses for, respectively, the vertical and the horizontal polarization. Given an element $a_{ij}$ of $\vec{A}$, we assume $\Re\{a_{ij}\},\Im\{a_{ij}\} \in C^0[\Omega]$.
\item $\alpha_c > 0$ is the average power of all the subpaths of cluster $c$, and it is assumed to be equal for UL and DL, which is a reasonable assumption for current FDD systems \cite{directional_properties}.
\item $\vec{B}: \Omega \to \mathbb{R}^{1\times2}$ is the frequency independent dual polarized antenna radiation pattern of the UE. The columns of $\vec{B}$ are denoted by $[b_V,b_H]:=\vec{B}$, and they represent, respectively, the radiation patterns for the vertical and for the horizontal polarization. We assume $\Re\{b_V\},\Im\{b_H\} \in C^0[\Omega]$.
\item The random matrix
\begin{equation*}
    \vec{M}_{ic} := 
\begin{bmatrix}
        e^{j\varphi_{VV,ic}}     & \dfrac{1}{\sqrt{K_i}}e^{j\varphi_{VH,ic}}       \\
         \dfrac{1}{\sqrt{K_i}}e^{j\varphi_{HV,ic}}     & e^{j\varphi_{HH,ic}}       \\
\end{bmatrix},
\end{equation*}
models the fading of the vertical and horizontal polarization, and also of the cross-polarization terms caused by the polarization changes that the electromagnetic waves undergo during the propagation. The components of the tuple $\bm{\varphi}_{ic}:=\{\varphi_{VV,ic},\varphi_{VH,ic},\varphi_{HV,ic},\varphi_{HH,ic}\}$ are i.i.d. random variables, uniformly distributed in $[-\pi,\pi]$. The UL and DL phases are assumed independent. The parameters $K_{ic} \in \mathbb{R}$, usually termed as cross polarization power ratios (XPRs), are assumed to be i.i.d. random variables and to be equal for UL and DL. This polarization propagation model is identical to the one suggested by \cite[Chapter~7]{MolishB}, where the two polarizations are assumed to experience independent fading.
\end{itemize}

We point out that, in contrast to \cite[Eq.~(7.3-22)]{3GPP-3D}, this model does not take into account the time dependent phase term $e^{j2\pi\nu_{ic}t}$, where $t$ is the time and $\nu_{ic}$ is the Doppler shift of subpath $i$ of cluster $c$, which models deterministically the short-term time evolution of the channel. However, as the focus of this work is on the long-term channel statistics, we consider only a long-term time evolution model, given in a statistical sense. More precisely, we model the time evolution of the channel as follows. The fast time-varying parameters $\bm{\theta}_{ic}$, $\bm{\phi}_{ic}$, $K_{ic}$ and $\bm{\varphi}_{ic}$ are drawn independently and kept fixed at intervals corresponding to the coherence time $T_c$ (``block-fading'' assumption). The slow time-varying parameters $\alpha_c$ and $f_c$ are assumed constant over a window $T_{WSS}$, with $T_{WSS} \gg T_c$. This model reflects the classical ``windowed WSS'' assumption, which approximates the channel as wide-sense stationary (WSS) for a  given time window $T_{WSS}$, which is usually several order of magnitude larger than $T_c$ \cite{Caire2,Cottat}. 

\subsection{Expression for the Spatial Covariance Matrix}
\label{ssec:covariance_expressions}
In the next proposition we present an expression for the spatial covariance matrices $\vec{R}^d$ and $\vec{R}^u$ on which the DL covariance estimation scheme proposed in Sect. \ref{sec:main_algo} is based.
\begin{proposition} By assuming the model introduced in Sect. \ref{ssec:realistic_multipath_model}, the spatial covariance matrices $\vec{R}^d:=\mathbb{E}\left[\vec{h}^d(\vec{h}^d)^H \right]$ and  $\vec{R}^u:=\mathbb{E}\left[\vec{h}^u(\vec{h}^u)^H \right]$ take the following forms:
\begin{equation}
\label{eq:Rd_expr}
\resizebox{1\columnwidth}{!}{$
    \vec{R}^d = \displaystyle \int_{\Omega}\rho_{V}(\bm{\theta})\vec{a}_{V}^d(\bm{\theta})\vec{a}_{V}^d(\bm{\theta})^Hd\bm{\theta} + \int_{\Omega}\rho_{H}(\bm{\theta})\vec{a}_{H}^d(\bm{\theta})\vec{a}_{H}^d(\bm{\theta})^Hd\bm{\theta}$} ,
\end{equation}
\begin{equation}
\label{eq:Ru_expr}
    \resizebox{1\columnwidth}{!}{$\vec{R}^u = \displaystyle \int_{\Omega}\rho_{V}(\bm{\theta})\vec{a}_{V}^u(\bm{\theta})\vec{a}_{V}^u(\bm{\theta})^Hd\bm{\theta} + \int_{\Omega}\rho_{H}(\bm{\theta})\vec{a}_{H}^u(\bm{\theta})\vec{a}_{H}^u(\bm{\theta})^Hd\bm{\theta}$} ,
\end{equation}
where the functions $\rho_V,\rho_H:\Omega \to \mathbb{R}^+$, referred to, respectively, as ``vertical polarization angular power spectrum'' (V-APS) and ``horizontal polarization angular power spectrum'' (H-APS), are defined to be
\begin{align*}
\begin{split}
    \rho_V(\bm{\theta}) &:= \sum_{c=1}^{N_c}\alpha_c\int_{\Omega}f_c(\bm{\theta},\bm{\phi})\left(b_V^2(\bm{\phi})+\dfrac{1}{K}b_H^2(\bm{\phi})\right)d\bm{\phi},\\
    \rho_H(\bm{\theta}) &:= \sum_{c=1}^{N_c}\alpha_c\int_{\Omega}f_c(\bm{\theta},\bm{\phi})\left(b_H^2(\bm{\phi})+\dfrac{1}{K}b_V^2(\bm{\phi})\right)d\bm{\phi}.\\
\end{split}
\end{align*}
Here $1/K:= \mathbb{E}[1/K_{ic}]$ is the average effect of the XPRs $K_{ic}$.
\end{proposition}

By recalling the notation defined in Sect. \ref{ssec:realistic_multipath_model}, we highlight that the V-APS and the H-APS do not depend on the carrier frequency. Furthermore, we have that $\rho_V,\rho_H \in L^2[\Omega]$.

\subsection{OFDM Systems}
\label{ssec:OFDM_expression}
We now show that expressions \eqref{eq:Rd_expr} and \eqref{eq:Ru_expr} (and hence the algorithms in Sect. \ref{sec:main_algo}) carry over to wide-band OFDM systems by extending the model in Sect. \ref{sec:model} with the approach in \cite{3GPP-3D} and \cite[Chapter~6]{MolishB} for the ``tapped delay line'' model. More precisely, we consider a wide-band channel in an under-spread environment; i.e., with delay spread $T_s \ll T_c$. By denoting with $l\in \mathbb{N}$ the discrete time index of the $l$th tap of the sampled impulse response, the channel vector $\tilde{\vec{h}}[k]$ in the sub-carrier domain is given by \cite[Chapter~3.4]{tse2005fundamentals}:
\begin{equation} 
\label{eq:OFDM_discrete}
\tilde{\vec{h}}[k] = \sum_{l=0}^{L-1}\vec{h}[l]e^{-j\dfrac{2\pi kl}{N_s}},
\quad \vec{h}[l] = \sum_{c=1}^{N_c}\vec{h}_c\delta[l - l_c],
\end{equation}
where $\{\vec{h}_c\}_{c = 1,\ldots,N_c}$ are defined in Sect. \ref{ssec:realistic_multipath_model}, $l_c\in \mathbb{N}$ denotes the discrete time delay of all the subpaths belonging to cluster $c$, $L$ is the impulse response length, $N_s$ is the chosen OFDM block length, and $k = 0, \ldots, (N_s-1)$ is the sub-carrier index. With this model in hand, we can derive expressions for the spatial covariance matrices in the sub-carrier domain. They are equivalent to the ones given by \eqref{eq:Rd_expr} and \eqref{eq:Ru_expr}, and they do not depend on the sub-carrier index. More precisely, we have:
\begin{proposition}
By assuming the wide-band OFDM channel model in \eqref{eq:OFDM_discrete}, the spatial covariance matrices $\vec{R}_k^d :=  \mathbb{E}\left[  \tilde{\vec{h}}^d[k] (\tilde{\vec{h}}^d[k])^H\right]$ and $\vec{R}_k^u := \mathbb{E}\left[  \tilde{\vec{h}}^u[k] (\tilde{\vec{h}}^u[k])^H\right] $ for a given sub-carrier $k$ satisfy 
\begin{equation*} 
\vec{R}_k^d  = \vec{R}^d, \quad
\vec{R}_k^u  = \vec{R}^u,
\end{equation*}
where $\vec{R}^d$ and $\vec{R^u}$ are given by \eqref{eq:Rd_expr} and \eqref{eq:Ru_expr}, and they do not depend on the sub-carrier index.
\end{proposition}

\section{Channel Spatial Covariance Conversion}
\label{sec:main_algo}

We now propose a practical FDD DL covariance estimation scheme based on the channel model described in Sect. \ref{sec:model}. The estimates of the DL channel covariance matrix $\vec{R}^d$ are obtained from the UL channel covariance matrix $\vec{R}^u$ by performing the following two-step scheme:
\begin{enumerate}[nosep]
\item Given $\vec{R^u}$, we obtain an estimate $(\hat\rho_V,\hat\rho_H)$ of $(\rho_V,\rho_H)$ from (\ref{eq:Ru_expr}) and known properties of $(\rho_V,\rho_H)$.
\item We compute the estimated $\vec{R^d}$ by using (\ref{eq:Rd_expr}) with $(\rho_V,\rho_H)$ replaced by their estimates $(\hat\rho_V,\hat\rho_H)$.
\end{enumerate} 
In this section, we assume perfect knowledge of $\vec{A}^u$, $\vec{A}^d$, and $\vec{R}^u$, while later in Sect. \ref{sec:UPA} and \ref{sec:sim} we assume that the BS have access only to noisy estimates of $\vec{R}^u$.

The core idea of the proposed scheme is that it is possible to address the joint V-APS and H-APS estimation problem of the first step as a \textit{convex feasibility problem}, which enables us to apply solutions based on projection methods. We point out that the related approaches in \cite{mypaper} cannot address properly the problem considered in this paper because they are based on a Hilbert space that is not appropriate to represent the \textit{estimandum} $(\rho_V,\rho_H)$ resulting from the channel model we consider here. 

To derive the proposed approaches, we first rewrite \eqref{eq:Ru_expr} as a system of equations of the form 
\begin{equation}
\label{eq:system_3Dpol}
    r_m^u = \int_{\Omega}\rho_V(\bm{\theta})g_{V,m}^u(\bm{\theta})d^2\bm{\theta} + \int_{\Omega}\rho_H(\bm{\theta})g_{H,m}^u(\bm{\theta})d^2\bm{\theta},
\end{equation}
where $r_m^u \in \mathbb{R}$ is the $m$th element of $ \setlength{\arraycolsep}{1.2pt}
\vec{r}^u := \text{vec}(
    \begin{bmatrix}
        \Re\{\vec{R}^u\} & \Im\{\vec{R}^u\}\\
    \end{bmatrix})
$, $g_{(\cdot),m}^u: \Omega \to \mathbb{R}$ is the corresponding $m$th coordinate function of \linebreak$\text{vec}(
    \begin{bmatrix}
        \Re\{\vec{a}_{(\cdot)}^u(\bm{\theta)}\vec{a}_{(\cdot)}^u(\bm{\theta})^H\} & \Im\{\vec{a}_{(\cdot)}^u(\bm{\theta)}\vec{a}_{(\cdot)}^u(\bm{\theta})^H\}\\
    \end{bmatrix}) $, and $ m = 1, \ldots, M$, with $M = 2N^2$. Now let $\mathcal{H}:=L^2[\Omega]\times L^2[\Omega]$ be the Hilbert space of tuples of bivariate square-integrable real functions equipped with the following inner product 
\begin{equation}\label{inner_product}
    \resizebox{1\columnwidth}{!}{$\langle (f_V,f_H),(g_V,g_H) \rangle := \displaystyle  \int_{\Omega}f_V(\bm{\theta})g_V(\bm{\theta})d\bm{\theta} +  \int_{\Omega}f_H(\bm{\theta})g_H(\bm{\theta})d\bm{\theta}$}.
\end{equation}
Based on the model in Sect. \ref{sec:model}, $(\rho_V,\rho_H)$ and $(g_{V,m}^u, g_{H,m}^u)$ are members of $\mathcal{H}$, thus \eqref{eq:system_3Dpol} can can be rewritten as
\begin{equation*}
    r_m^u = \langle(\rho_V,\rho_H),(g_{V,m}^u, g_{H,m}^u) \rangle \quad m = 1, \ldots, M.
\end{equation*}
By using the set-theoretic paradigm \cite{SetTheo1,SetTheo2,SetTheo3,SetTheo4}, we obtain an estimate $(\hat\rho_V,\hat\rho_H)$ of $(\rho_V,\rho_H)$ by solving one of the two following feasibility problems:
\begin{equation}\label{eq:conv_feas_1}
   \text{find } (\hat\rho_V,\hat\rho_H) \in V:= \cap_{m=1}^M V_m \neq \emptyset,
\end{equation}
\begin{equation}\label{eq:conv_feas_2}
   \text{find } (\hat\rho_V,\hat\rho_H) \in C:= V \cap Z \neq \emptyset,
\end{equation}
where $V_m := \{(h_V,h_H) \in \mathcal{H}: \langle (h_V,h_H), (g_{V,m}^u,g_{H,m}^u) \rangle = r_m^u \}$ are hyperplanes and $Z := \{(h_V,h_H) \in \mathcal{H}: (\forall \bm{\theta} \in \Omega) \quad h_V(\bm{\theta})\geq 0, h_H(\bm{\theta})\geq 0\}$ is the cone of tuples of non-negative functions. We solve problem \eqref{eq:conv_feas_1} by computing the projection onto the linear variety $V$, while problem \eqref{eq:conv_feas_2}, which takes into account also the positivity of $\rho_V$ and $\rho_H$, is solved via an iterative projection method called \textit{extrapolated alternating projection method (EAPM)}. For the details about the solutions of the considered feasibility problems, we refer to \cite{mypaper} and to the references herein. 

The choice of solving either \eqref{eq:conv_feas_1} or \eqref{eq:conv_feas_2} leads to two variants of the proposed scheme with different complexity and accuracy, and they are referred here as \textit{Algorithm 1} and \textit{Algorithm 2}. More precisely, \textit{Algorithm 1} can be implemented as a simple matrix multiplication of the form $\vec{r}^d = \vec{F}\vec{r}^u$, where $\vec{F}$ depends just on the array geometry and can be computed once for the entire system lifetime. In contrast, \textit{Algorithm 2} requires iteratively the evaluation of integrals of the form $\int_{\Omega}x(\bm{\theta})d^2\bm{\theta}$ (see \cite{mypaper} for details).

\section{Implementation for Uniform Planar Array with Pairs of Cross-Polarized Antennas}
\label{sec:UPA}

In this section we describe implementation aspects for a cross-polarized uniform planar array (UPA), defined here as a rectangular grid of identical and equispaced antenna elements, each of them composed of a pair of two vertically polarized antennas with a polarization slant of $\pm 45^\circ$. We denote by $N_V$ and $N_H$ the number of vertical and horizontal elements, respectively, and by $d$ the inter-antenna spacing. We further denote by $x(u,v,1)$ the antenna in position $(u,v)$, $u = 1, \ldots ,N_V$ and $v = 1, \ldots ,N_H$, with $+45^\circ$ polarization slant, and by $u(u,v,2)$ the co-located antenna with $-45^\circ$ polarization slant. For this antenna array, the covariance matrix has the following structure:

\begin{proposition}[Structure of the UPA Covariance Matrix]
\label{prop:UPA_struct}
By letting $
    \vec{h} := \begin{bmatrix}
        \vec{h}_1^T & \vec{h}_2^T \\
    \end{bmatrix}^T,
$
where the channel coefficient for antenna $x(u,v,k)$ corresponds to the $n$th element of the vector $\vec{h}_k\in\mathbf{C}^{N_VN_H\times 1}$, with $n = (u-1)N_H + v$, and by assuming without loss of generality that $N_V \geq N_H$, the covariance matrix takes on the following block structure:
\begin{equation*}
    \vec{R} = \begin{bmatrix}
            \vec{B}_1 &  \vec{B}_2^H\\
            \vec{B}_2 &  \vec{B}_3 \\
    \end{bmatrix} \in\mathbf{C}^{2N_VN_H\times 2N_VN_H},
\end{equation*}
where every macro-block $\vec{B}_l \in\mathbf{C}^{N_VN_H\times N_VN_H}$, $l = 1,2,3$, is Hermitian and it has the following block structure:
\begin{equation*}
    \vec{B}_l = \begin{bmatrix}
            \vec{B}_{l,1}   &                &               &               &               \\
            \vec{B}_{l,2}   &  \vec{B}_{l,1} &               &               &               \\
            \vec{B}_{l,3}   &  \vec{B}_{l,2} & \vec{B}_{l,1} &               &               \\
            \vdots          &  \vdots        & \vdots        & \ddots        &               \\
            \vec{B}_{l,N_V} &  \ldots        & \vec{B}_{l,3} & \vec{B}_{l,2} & \vec{B}_{l,1} \\
    \end{bmatrix},
\end{equation*}
where every block $\vec{B}_{l,i} \in\mathbf{C}^{N_H\times N_H}$, $i=1, \ldots N_V$ has identical diagonal entries $b_{li}$, and every block $\vec{B}_{l,1}$ is Hermitian Toeplitz. 
\end{proposition}
The proof is omitted here but we point out that it follows by direct inspection of the matrices $\vec{a}_V(\bm{\theta)}\vec{a}_V(\bm{\theta})^H$ and $\vec{a}_H(\bm{\theta)}\vec{a}_H(\bm{\theta})^H$ of \eqref{eq:Rd_expr} and \eqref{eq:Ru_expr}, where the elements of the array responses $\vec{a}_V$ and $\vec{a}_H$ are arranged with the same scheme adopted for $\vec{h}$.

The structure of the UPA covariance matrix $\vec{R}$ described in Prop. \ref{prop:UPA_struct} has the following consequences in practical implementations of the algorithms presented in Section \ref{sec:main_algo}:
\begin{itemize}[topsep=0pt,parsep=0pt,partopsep=0pt,leftmargin=10pt,labelwidth=6pt,labelsep=4pt]
    \item $\vec{R}$ can be bijectively vectorized by using only $M = 6(N_H + (N_V -1)(N^2_H - N_H +1))$ real numbers, compared to the $M = 2(N_VN_H)^2$ elements given by the vectorization operation defined in Sect. \ref{sec:main_algo}. 
    \item Any estimate $\hat{\vec{R}}$ of the covariance matrix $\vec{R}^u$ (for example, obtained from the sample covariance matrix similarly to \cite[Sect. ~4.2.]{mypaper}) can be further improved by substituting each element $(i,j)$ with the arithmetic average of all the elements that are assumed to be identical. 
\end{itemize}

\section{Simulation}
\label{sec:sim}

In this section we evaluate the proposed algorithms by simulating a communication scenario with system parameters given in Table \ref{tab:general_param}. The channel coefficients are given by the narrow-band multipath model described in Sect. \ref{ssec:realistic_multipath_model}, with parameters randomly drawn as follows:

\begin{table}[!t]
    \caption{General simulation parameters}
    \label{tab:general_param}
    \centering
    \begin{tabular}[h]{c|c}
        \hline
        Carrier frequency ($f_c$) & 1.8 GHz for UL, 1.9 GHz for DL            \\
        System type & Narrow-band or wide-band OFDM                           \\
        BS         & 8x4 cross-polarized UPA                                 \\
                    & $d = \lambda_u/2$                               \\
        UE         & Single antenna, vertically polarized                    \\
        Antennas radiation pattern & 3GPP \cite[Section~7.1]{3GPP-3D}, 3D-UMa \\
        \hline
    \end{tabular}
		\vspace{-0.3cm}
\end{table}

\begin{itemize}[topsep=0pt,parsep=0pt,partopsep=0pt,leftmargin=10pt,labelwidth=6pt,labelsep=4pt]
    \item Cluster powers $\alpha_c$ are drawn uniformly from $[0,1]$ and further normalized such that $\sum_{c=1}^{N_c}\alpha_c = 1$. 
    \item The XPRs values $K_{ic}$ are drawn from a log-Normal distribution with parameters $(\mu_{\text{XPR}}, \sigma_{\text{XPR}}) = (7,3)[\text{dB}]$. This is identical to the 3GPP model \cite[Sect. ~7.3, Step 9]{3GPP-3D}, with parameters for 3D-UMa, NLOS propagation.
    \item The angles $\bm{\theta}_{ic}$, $\bm{\phi}_{ic}$ are generated from the jointly Gaussian distribution $f_c(\bm{\theta},\bm{\phi}) =   f_{\text{BS},c}(\bm{\theta})f_{\text{UE},c}(\bm{\phi})$, where $f_{\text{BS},c}\sim \mathcal{N}(\bm{\mu}_{\text{BS}},\bm{\sigma }^2_{\text{BS}}\vec{I})$ and $ f_{\text{UE},c}\sim \mathcal{N}(\bm{\mu}_{\text{UE}},\bm{\sigma }^2_{\text{UE}}\vec{I})$, and where the clusters means and angular spreads 
    \begin{align*}
       \bm{\mu}_{\text{BS}} := [\mu_{\text{BS},a}\quad\mu_{\text{BS},z}], &\quad
       \bm{\sigma }^2_{\text{BS}} := [\sigma^2_{\text{BS},a} \quad\sigma^2_{\text{BS},z}],\\
       \bm{\mu}_{\text{UE}} := [\mu_{\text{UE},a}\quad\mu_{\text{UE},z}], &\quad
       \bm{\sigma }^2_{\text{UE}} := [\sigma^2_{\text{UE},a} \quad \sigma^2_{\text{UE},z}],
    \end{align*}
    are drawn as follows:\\$\mu_{\text{BTS},a},\mu_{\text{UE},a}$ are uniformly drawn from$\left[-\frac{2\pi}{3},\frac{2\pi}{3}\right]$,  $\mu_{\text{BTS},z},\mu_{\text{UE},z}$ from $\left[\frac{\pi}{4},\frac{3\pi}{4}\right] $, $ \sigma_{\text{BTS},a}$ from $\left[3^\circ,5^\circ\right]$, $ \sigma_{\text{UE},a} $ from $\left[5^\circ,10^\circ\right]$,  $\sigma_{\text{BTS},z}$ from $\left[1^\circ,3^\circ\right]$, and $ \sigma_{\text{UE},z}$ from $\left[3^\circ,5^\circ\right]$. This choice of parameters is inspired by experimental properties of $\rho_V$ and $\rho_H$ given by \cite{3GPP-3D}, e.g. the elevation angular spread is usually narrower than the azimuth one. 
    \item To simulate different UE antenna orientation, the UE antenna array response is given by applying a 3D rotation to the antenna radiation pattern  as described in 3GPP \cite[Sect.~5.1.3]{3GPP-3D}, with parameters
    $\alpha, \beta, \gamma \sim \mathcal{U}\left[0, \frac{\pi}{6}\right]$.
\end{itemize}

The BS is assumed to have access to the estimated UL covariance matrix $\hat{\vec{R}}^u$ obtained from $N_s = 1000$ noisy channel estimates $\hat{\vec{h}}^u = \vec{h}^u + \vec{z}$, $\vec{z} \sim \mathcal{C}\mathcal{N}(\vec{0}$, $\sigma^2_z\vec{I})$ i.i.d., with noise power defined by setting an average per-antenna $\text{SNR}_{est}$ to $\text{SNR}_{est} := \text{tr}\{\vec{R}^u\}/(N\sigma_z^2) = 10$ [dB], where $N = 2N_VN_H$ denotes the number of BS antennas. The estimate $\hat{\vec{R}}^u$ is computed by projecting the sample covariance matrix onto the space of positive semi-definite matrix $\mathcal{C}$ as described in \cite[Sect.~4.2]{mypaper}, and by applying the correction procedure described in Sect. \ref{sec:UPA}. Furthermore, the proposed algorithms are implemented by exploiting the efficient vectorization for UPA described in Sect. \ref{sec:UPA}.

The accuracy of an estimate $\hat{\vec{R}}^d$ of $\vec{R}^d$ is evaluated in terms of the square error $\text{SE} := e^2(\vec{R}^d,\hat{\vec{R}}^d)$, where $e(\cdot,\cdot)$ is a given error metric. In particular, we consider as error metrics the normalized Frobenius norm and the 90\% Grassmanian principal subspace distance defined in \cite[Sect.~5]{mypaper}.

\begin{figure} 
    \centering
  \subfloat[Normalized Euclidean distance]{
       \includegraphics[trim=0.5cm 0.15cm 0.5cm 0.9cm, clip,width=0.75\linewidth]{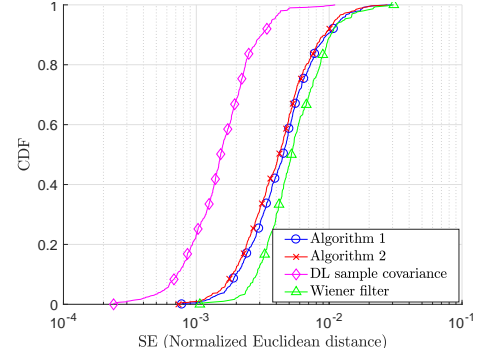}}
    \label{fig:perf_frob}\hfill \\
  \subfloat[Principal subspaces distance]{
        \includegraphics[trim=0.5cm 0.1cm 0.5cm 0.9cm, clip,width=0.75\linewidth]{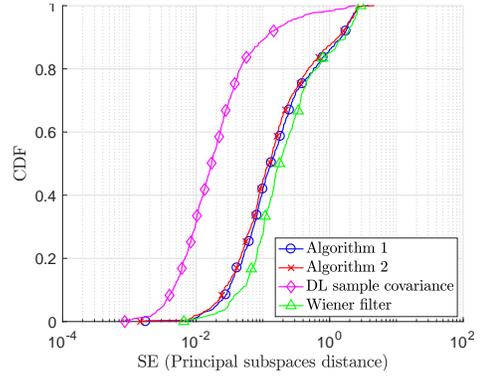}}
    \label{fig:perf_subs}\\
  \caption{Empirical CDF of the squared error (SE)}
  \label{fig_sim} 
  \vspace{-0.5cm}
\end{figure}

To evaluate the proposed algorithms we use as a baseline an estimate of the DL covariance matrix obtained from $\hat{\bm{h}^d}$ samples with the same technique for the estimation of $\hat{\vec{R}}^u$. Furthermore, we also compare the proposed approaches with a solution that relies on a pre-stored dictionary of covariance matrices $(\hat{\vec{R}}^u,\hat{\vec{R}}^d)$ and based on the Wiener filter, similar to the approach proposed in \cite{Dict1} that was already analyzed with the preliminary results in \cite[Sect. ~5]{mypaper}. The results are shown in Figure \ref{fig_sim}, which shows the empirical cumulative distribution function (CDF) of the SE for the two chosen metrics, obtained by drawing independent realizations of the quantities that are assumed to stay fixed for $T_{\text{WSS}}$ (i.e. by drawing a new V-APS and H-APS). The simulation confirms that the proposed algorithms are able to provide an accurate DL estimate by using only UL training, thus it can be used as an effective solution to the DL channel covariance acquisition problem. With respect to \textit{Algorithm 1}, \textit{Algorithm 2} is slightly more accurate, as it considers also the non-negativity property of the V-APS and H-APS, but it pays a price in terms of increased complexity. We point out that, as opposed to the Wiener filter approach and to all other techniques based on supervised machine learning tools, the two proposed algorithms do not require any training phase.

\bibliographystyle{IEEEbib}
\bibliography{refs}

\end{document}